\documentstyle[twoside,epsf,12pt]{kaeog}

\newfont{\mm}   {msbm10}              
\newcount\bighand\newcount\littlehand
\bighand=\time\divide\bighand by 60
\littlehand=\bighand\multiply\littlehand by -60
\advance\littlehand by\time
\def\timenow{~\the\bighand:\ifnum\littlehand<10{0}\fi\the\littlehand}

\begin{document}

\author{Marek W. Gutowski\\
\medskip
	{\small \sl Institute of Physics,
	Polish Academy of Sciences\\
	02--668 Warszawa, Al. Lotnik\'ow 32/46,
	Poland\\
	e-mail: gutow@ifpan.edu.pl}
}

\title{\Large\sf \uppercase{Power and beauty of interval methods}}


\maketitle


\begin{abstract}
Interval calculus is a relatively new branch of mathematics.
Initially understood as a set of tools to assess the quality of
numerical calculations (rigorous control of rounding errors), it became
a discipline in its own rights today. Interval methods are useful
whenever we have to deal with uncertainties, which can be rigorously
bounded.  Fuzzy sets, rough sets and probability calculus can perform
similar tasks, yet only the interval methods are able to (dis)prove,
with mathematical rigor, the (non)existence of desired solution(s). 
Known are several problems, not presented here, which cannot be
effectively solved by any other means.

This paper presents basic notions and main ideas of interval calculus
and two examples of useful algorithms.

\end{abstract}



\section{What is an interval anyway?}
\noindent
{\bf Definition: } {\sl The {\bf interval} is a bounded subset of real
numbers.  Formally:}
$$
\bigl(X=\left[a,b\right]\ \hbox{\rm is an {\bf interval}}\bigr) \iff
\bigl( X=\{x\in\hbox{\mm R} \vert\ a\le x\le b\}\bigr),
$$
{\sl where $a$, $b \in\hbox{\mm R}$ (set of all real numbers); in particular
$a$, $b$, or even both of them may be infinite.}

\medskip\noindent
Geometrically, interval is just a section of a real line, uniquely
determined by its own endpoints.  The set of all intervals is commonly
denoted as \hbox{\mm IR}.  Lower and upper endpoint of interval $X$ is
usually referred to as $\underline{X}$ and $\overline{X}$,
respectively.  Intervals with property $\underline{X}=\overline{X}$ are
called {\bf thin} (or {\em degenerate\/}), any of them contains exactly
one real number and can be thus formally identified with this very
number.  Two basic real-valued functions defined on intervals (i.e. of
type $\hbox{\mm IR}\mapsto\hbox{\mm R}$) are: {\bf width}: ${\rm
w}\,(X)=\vert X\vert\buildrel{\rm def}\over{=}
\vert\overline{X}-\underline{X}\vert$,\ and {\bf center}:
${\rm mid}\,(X)\buildrel{\rm def}\over{=}\frac{1}{2} \vert
\underline{X} + \overline{X}\vert$.

\medskip\noindent
{\bf Algebraic operations on intervals} are defined in such a way that
their results always contain {\bf every possible} outcome of the
corresponding algebraic operation on real numbers.  More specifically:
the result of $X\diamond Y$ is again an interval, $Z$, with
property
$$
	X\diamond Y=Z=\{z=x\diamond y\ \vert\ x\in
	X,\ y\in Y\},
$$
where $\diamond$ belongs to the set $\{+, -, \times, \slash\}$.  One can
easily prove that arithmetic operations on intervals can be expressed in
terms of ordinary arithmetics on their endpoints:
$$
	X + Y = \left[ \underline{X} + \underline{Y},\
	\overline{X} + \overline{Y}\right]\quad\quad\quad\
	X - Y = \left[\underline{X} - \overline{Y},\
	\overline{X} - \underline{Y}\right]
$$
$$
	X\cdot Y = \left[
	\min\,(\underline{X}\cdot\underline{Y},\
	\underline{X}\cdot\overline{Y},\
	\overline{X}\cdot\underline{Y},\
	\overline{X}\cdot\overline{Y}),\
	\max\,(\underline{X}\cdot\underline{Y},\
	\underline{X}\cdot\overline{Y},\
	\overline{X}\cdot\underline{Y},\
	\overline{X}\cdot\overline{Y})\right]
$$
$$
	X/Y = \left[
	\min\,(\underline{X}/\underline{Y},\
	\underline{X}/\overline{Y},\
	\overline{X}/\underline{Y},\
	\overline{X}/\overline{Y}),\
	\max\,(\underline{X}/\underline{Y},\
	\underline{X}/\overline{Y},\
	\overline{X}/\underline{Y},\
	\overline{X}/\overline{Y})\right]
$$
with an extra condition for division: $0\notin Y$.

\medskip\noindent
In computer realization we have to take care of proper rounding of
every intermediate result in order to preserve the property that the
final results are guaranteed.  The appropriate rounding is called {\em
outward\/} or {\em directed rounding\/}, i.e. $\underline{Y}$ must be
always rounded down (`towards $-\infty$') and $\overline{Y}$ has to be
rounded up (`towards $+\infty$').  This is achieved either in hardware,
by proper switching back and forth the processor's rounding mode (still
rare), or in software as {\em simulated rounding\/} (majority of
existing software packages).

\medskip\noindent
{\bf Intervals as sets.}  Since intervals are sets, it is possible to
carry typical set operations on them. For example we can consider the
{\bf intersection}s of intervals, like $Y=X_{1}\cap X_{2}$.  However,
the intersection of two disjoint intervals is an empty set!  This shows
the necessity of considering the {\bf empty interval} as a legitimate
member of the set {\mm IR}. It is usually denoted as $\emptyset$ and in
machine representation should be, for many reasons, expressed as
$\left[INF, -INF\right]$, where $INF$ is the largest
machine-representable positive number.

\medskip\noindent
Unfortunately, the {\bf union} of two intervals is not always an
interval.  Instead, we can define the {\bf interval hull} of two
arbitrary intervals (or of any other subset of {\mm R} as well) as the
smallest interval containing them both:
$$
	{\rm hull}\,(X, Y) = \left[\right](X, Y) =
	X\underline{\cup}Y \buildrel{\rm def}\over{=}
	\left[\min\,(\underline{X},\ \underline{Y}),\
	\max\,(\overline{X},\ \overline{Y}) \right]
$$
There is no problem with checking whether $X\subset Y$ (or $X\subseteq Y$).

\medskip\noindent
It is worth to mention that addition and multiplication of intervals
are both commutative: $X\diamond Y=Y\diamond X$, and associative:
$X\diamond Y\diamond Z = (X\diamond Y)\diamond Z=X\diamond(Y\diamond
Z)$.  However, it is surprising that in general the following holds:
$$
X\cdot (X+Z) \subseteq X\cdot Z + X\cdot Y
$$
(and {\em not\/} just equality!) i.e. the multiplication is only {\em
subdistributive\/} with respect to addition.  We can also see, that
using the same variable more than once (here: $X$), in rational
expression, leads inadvertently to the overestimation of the final
result.  This phenomenon is known under the name of {\bf dependency
problem}. This pro\-per\-ty, together with the lack of good order in
{\mm IR} ({\mm IR} is only partially ordered set ({\em poset\/})) makes
conversion of ordinary computer programs into their interval
equivalents not a straightforward task.  In particular, every computer
instruction of the type
\vfill
\begin{quote}
{\bf if} $x<y$\\
\phantom{\bf ifxxx}{\bf then} \ldots\\
\phantom{\bf ifxxx}{\bf else} \ldots\\
{\bf end if}
\end{quote}
\vfill
has to be carefully redesigned.

\medskip\noindent
{\small
Example:
\begin{quote}
Before shopping I had exactly $100.00$ monetary units in my pocket and
my wife had something between $42.00$ and $45.00$ units. So we had
together $\left[142,145\right]$ units.  We have already spent $127.99$
units, so we still have $\left[14.01,17.01\right]$ units. Can we afford
to call a taxi with estimated cost $\left[13,15\right]$ units?  {\em
Possibly\/} \dots\\
But, without any credit,  we {\em certainly\/} can buy two bottles of
milk at the cost
$2\times\left[1.19,3.59\right]=\left[2.38,7.18\right]$ units.
Guaranteed.
\end{quote}
}
\medskip\noindent
{\bf Interval vectors and matrices.} Any $n$-dimensional vector with at
least one interval component may be called interval vector or {\bf
box\/}. Any matrix with at least one interval entry will be called an
interval matrix.  Ordinary linear algebra can be done on those objects,
if only every elementary arithmetic operation is substituted by its
suitable interval counterpart, as defined earlier. The most often used
norms for interval vectors are: $\Vert\cdot\Vert_{1}$, equal to the sum
of widths of all its components, and $\Vert\cdot\Vert_{\infty}$ being
the width of the widest component.

\section{Interval functions}
\noindent
An obvious requirement for the good interval substitute $F$ of the
real-valued function $f$ is following
$$
	F(X) = \left[\right]\mathop{f(x)}_{x\in X}=\left[\inf_{x\in X}
	f(x),\ \sup_{x\in X}f(x)\right]
$$
We would call such an $F(X)$ a {\bf range function} for $f$.  The
explicit construction of range function may be difficult, so we often
work with the so called {\bf inclusion} or {\bf inclusive functions}. 
These are not unique, but any such function satisfies
$$
	F(X)\supseteq \mathop{f(x)}\quad\forall{x\in X}
$$
$F$ is also called an {\bf interval extension} for $f$.
Note, that:
\begin{itemize}
\item $F$ may be `broader' than the range function, i.e. it usually
overestimates the range of function $f$, and
\item there is no explicit specification how large (or small) this
overestimation can be.
\end{itemize}

\medskip\noindent
The most desirable are the so called {\bf monotonic inclusion
function}s, i.e. such inclusion functions, which additionally satisfy
the implication
$$
	\biggl(X\to x\biggr)\Rightarrow \biggl(F(X)\to
	f(x)\biggr)\quad\ \forall{x\in X}
$$
more properly formulated as
$$
	\biggl(X_{1}\subset X_{2}\biggr)\Rightarrow\biggl(F(X_{1})
	\subset F(X_{2})\biggr)
$$
This is only possible for functions, which are everywhere continuous.
Shortly one can say that range functions and monotonically inclusive
functions produce thin intervals for thin arguments, while functions,
which are only inclusive generally return `true' (i.e. non-degenerate)
intervals --- even if their arguments are thin.

\medskip\noindent
{\small
Example:
\begin{quote}
Let $f(x)={\rm sign}\,x=\cases{+1 & when $x>0$ \cr
				\phantom{+}0 & when $x=0$ \cr
				-1 & when $x<0$
}$

\medskip\noindent
The range function corresponding to sign is:
$${\rm SIGN}\,(X) =\cases{ \phantom{[}+1 & when\ \ $\underline{X}>0$ \cr
		\left[\phantom{+}0,+1\right] & when\ \ $\underline{X}=0$ \cr
		\left[-1,+1\right] & when\ \ $\underline{X}\overline{X}<0$ \cr
		\left[-1,\phantom{+}0\right] & when\ \ $\overline{X}=0$ \cr
		\phantom{[}-1 & when\ \ $\overline{X}<0$ \cr
		0 & when\ \ $X=\left[0,0\right]$ }
$$
and one of its many inclusion functions may be given as
$$
F(X)=\left[-1.5,\ 2.5\right]
$$
while no monotonically inclusive function exists for this case, since
the original function is discontinuous.
\end{quote}
}
\medskip\noindent
{\bf Important remark.} The value of the range function for argument
$X$ should be calculated only for $x\in X\cap{\mathcal
D}\left(f\right)$, where ${\mathcal D}(f)$ is domain of $f$.  Empty set
should be returned whenever $X\cap{\mathcal D}(f)=\emptyset$. 
Therefore $\sqrt{\left[-4,+4\right]}=\left[0,2\right]$ and
$\sqrt{\left[-20,-10\right]}=\emptyset$.

\section{Interval-oriented algorithms}
\noindent
As George Corliss pointed out, usual (i.e. non-interval) algorithms
only rarely are a good starting point for interval oriented ones.  The
vast majority of work done so far was concentrated on optimization
problems and on solving systems of algebraic equations in many
variables.  There are remarkable results achieved in this field with
interval version of Newton method being the most honored.

\medskip\noindent
The typical example of interval methods is the algorithm due to Ramon
E.~Moore and Stieg Skelboe, which belongs to the class of `divide and
bound' algorithms.  Suppose our task is to find the global minimum of a
real-valued function $f$ of $n$ variables over the box
$V_{0}=X_{1}\times X_{2}\times\cdots\times X_{n}$.  The initial step is
to construct an interval extension $F$ for the function $f$. The
algorithm operates on the list of $n$-dimensional boxes, ${\mathcal
L}$, which initially contains the only element, a pair: the box $V_{0}$
and the interval $F(V_{0})$. We will also need a real number,
$f_{test}$, initially equal to $\uparrow\!\!f({\rm any}\ x\in V_{0})$
or just $\overline{F}(V_{0})$. The outline of the rest of algorithm, in
pseudo code, follows:

\vfill
\begin{quote}
{\bf do while} {\sl diameter of the first box on the list ${\mathcal L}$ exceeds
some predefined value}\\
\hbox{\hspace{2em}} {\sl pick the first element $V$ and its bounds $F(V)$ from list ${\mathcal
	L}$}\\
\hbox{\hspace{2em}} {\sl remove $V$ from} ${\mathcal L}$\\
\hbox{\hspace{2em}} {\bf if}\ $\underline{F}(V)\le f_{test}$\ {\bf then}\\
\hbox{\hspace{4em}} {\sl bisect $V$ perpendicularly to its longest edge
	obtaining} $V_{1}\cup V_{2}=V$\\
\hbox{\hspace{4em}} {\sl calculate intervals $F_{1}=F(V_{1})$ and $F_{2}=F(V_{2})$}\\
\hbox{\hspace{4em}} {\bf for} $i=1,2$\ {\bf do}\\ 
\hbox{\hspace{6em}} {\bf if}\ $\underline{F_{i}} > f_{test}$\\
\hbox{\hspace{6em}} {\bf then} {\sl discard box} $V_{i}$\\
\hbox{\hspace{6em}} {\bf else} {\sl put pair $(V_{i}, F_{i}(V_{i}))$ at the end of ${\mathcal
	L}$}\\
\hbox{\hspace{8em}} $f_{test}\leftarrow\min\,\left(f_{test},\
	\overline{F_{i}},\ \uparrow\!\!f({\rm center\ of}\ V_{i})\right)$\\
\hbox{\hspace{6em}} {\bf end if}\\
\hbox{\hspace{4em}} {\bf end for}\\
\hbox{\hspace{2em}} {\bf end if}\\
{\bf end do}
\end{quote}
\vfill
The operation `$\uparrow$' means {\em round the next number up\/}.  The
algorithm continuously `grinds' boxes on the list ${\mathcal L}$,
making them smaller and smaller.  Some of them disappear forever.  At
exit we can say that the global minimizer(s) $x^{\star}$, such that
$f(x^{\star})=\min_{x\in V_{0}}f(x)$, is (are) contained with certainty in
the union of all the boxes still present on ${\mathcal L}$.  Numerous
variants of the above algorithm do exist for less general cases, for
example, when $f$ is differentiable almost everywhere in $V_{0}$.  It
is absolutely essential, from the performance point of view, to get rid
of `bad' (sub)boxes as early as possible. And the reason is clear: to
test all subboxes, which are twice smaller than $V_{0}$ (in each
direction), it is necessary to consider up to $2^n$ of them.  The
properties of $f$ and its interval extension $F$ as well, can influence
the speed of convergence, which may be arbitrarily slow.

\medskip\noindent
Due to space limitations, we have to stop here with this introductory
course.  More, and most likely better, materials can be found in the
web \cite{Intro}.  The excellent starting point, with pointers to other
valuable sites, is also \cite{ftp}.  Those, who prefer classical forms
are encouraged to see the book \cite{franc}.

\section{Where are we today?}
\noindent
Interval analysis started as a part of numerical analysis, devoted
mainly to automatic verification of computer-generated results.  The
four basic arithmetic operations were everything what was needed for
this purpose.  There were two goals in front of researchers and users
of interval calculus:
\begin{itemize}
\item to obtain guaranteed bounds for results in every case, and
\item to make every possible effort to have those bounds as tight
as possible.
\end{itemize}
They are still important, therefore better and better methods for
construction of inclusion functions are discussed.  Besides naive
(natural) expressions we have at our disposal mean value theorem,
Lipschitz forms, centered forms, and --- recently --- Taylor centered
forms.  After (re)discovering various old theorems, and proving new
ones, it became clear, that interval methods, mostly those based on
fixed point theorems, have enormous power to prove or disprove, with
mathematical rigor, the existence of solutions to nonlinear systems of
equations.  As a complete surprise we learned that some problems,
thought hopeless, can be successfully attacked with interval methods,
while no other method apply.

\medskip\noindent
Two kinds of research activity is visible today:
\begin{itemize}
\item introduction of interval methods into other branches of `hard'
science, like physics, astronomy or chemistry, as well as into
engineering and business everyday practice
\item establishing connections with other branches of pure and applied
mathematics like, for example, fuzzy set theory, mathematical
statistics and others.
\end{itemize}

\medskip\noindent
The first area is `easy'.  Just learn, implement and use.  Continuously
increasing computing power makes interval calculations feasible and
acceptable, regardless that they are usually 8--20 times slower than
their regular floating-point counterparts.  This is no longer a serious
problem.  Commercial and free software is also easily available.

\medskip\noindent
The second kind of activity goes much deeper.  New ideas are emerging,
interval methods inspire specialists from other fields.  One can
clearly notice gradual shift of interest into, generally speaking,
imprecise probability theory.  Practical consequences are important in
environmental protection, risk analysis, robotics, fuzzy sets theory
and applications, experimental data processing, quality control,
electric power distribution, constraint propagation, logic programming,
differential equations --- to name a few.

\section{New paradigms in experimental sciences}
\noindent
{\em Parameter identification\/} in engineering and {\em data
fitting\/} in experimental sciences are code words for nearly the same
thing.  The task of reconstruction of values of unknown
pa\-ra\-me\-ters, given experimental observations, lies at the heart of
the so called {\em inverse problems\/}.  The problem is usually
formulated as follows:
\vfill
\begin{quote}
{\bf given}:
\begin{itemize}
\item $N$ observations $y_1, y_2, \ldots, y_N$,
\item taken for the corresponding values $x_1, x_2, \ldots, x_N$ of the
control variable $x$,
\item depending additionally on $p$ unknown parameters $a_1, a_2,
\ldots, a_p,\quad p<N$ 
\item and the mathematical model, $f(x,y,{\mathbf a})=0$, relating $y$'s
with $x$'s and with the constant vector ${\mathbf a}$
\end{itemize}
{\bf find} the numerical values of all parameters $a_1, a_2, \ldots, a_p$.
\end{quote}

\medskip\noindent
There is a bunch of, more or less standard, approaches to this problem,
especially, when the relation $f(x,y,{\mathbf a})=0$ is simply
a~function $y=f(x,{\mathbf a})$.  The most popular are: least squares
method (LSQ), least absolute deviations (LAD) and maximum entropy
methods (MEM).  All they are based on finding the absolute (global)
minimum of the appropriately chosen functional.  We would like to find
the most appropriate set of unknown parameters, which is also the
minimizer of such a~functional.  It is obvious, that the final result
may vary, depending on which functional shall be used.

\medskip\noindent
Let us now present the interval-type approach to this very problem.  We
will replace minimization procedure by solution of suitable constraint
satisfaction problem. Both the $x$'s and $y$'s, due to unavoidable
experimental uncertainties, should be treated as intervals containing
the (unknown) true value.  We will assume, that those intervals are
indeed guaranteed, i.e. they contain the true values of control
variables, and measured results respectively, with probability equal
exactly to $1$.  We will search {\em not\/} for the most likely values
of unknown parameters ${\mathbf a}$, but for their {\em possible\/}
values instead.  For example, when fitting the straight line (extension
for more complicated cases is immediate) $y=a\,x+b$, (parameters
${\mathbf a}=\left(a, b\right)$), we will consider the set of
relations:
$$
\left\{ \left(a\,x_j + b\right) \cap y_j \ne \emptyset\quad\ j=1, 2,
\ldots, N\right.
$$
In geometrical interpretation the above means that the straight line
with slope in the interval $a$ and intercept in the interval $b$, both
yet unknown, passes through every `uncertainty rectangle' $x_i\times
y_i$, $i=1, 2, \ldots\, N$.  In purely algebraic terms:
\begin{equation}\label{ineq_sys}
\bigl(\left(a\,x_j + b\right) \cap y_j \ne \emptyset\bigr) \iff \bigl(
\underline{a\,x_j+b} \le \overline{y}_j\ \land\
\overline{a\,x_j+b} \ge \underline{y}_j\bigr)
\end{equation}
This way {\bf the data} themselves and {\bf their uncertainties}, with
{\bf no additional assumptions}, determine the intervals for possible
values of unknown parameters $a$ and $b$.  Such a possibility was first
pointed out by Walster \cite{Bill} in 1988.  To discover the intervals
$a$ and $b$ we will use the following procedure:
\vfill
\begin{quote}
\begin{enumerate}
\item {\sl start with initial box $V=\left(a,b\right)$ such that all
inequalities (\ref{ineq_sys}) are\/ {\bf possibly} satisfied somewhere
within $V$ but\/ {\bf certainly} not on their faces.}
\item\label{repeat} {\sl working with $V^{\prime}$, the exact copy of\ $V$,
and using\/ {\bf box slicing} algorithm, obtain its new version taking
into account all the inequalities\ $\strut{\underline{a\,x_j+b}}\le
\strut{\overline{y}_j}$\ only.}
\item {\sl working with $V^{\prime\prime}$, another exact copy of\ $V$, and
using\/ {\bf box slicing} algorithm again, obtain its new version when
only the inequalities\
$\strut{\overline{a\,x_j+b}}\ge\strut{\underline{y}_j}$\ are all
satisfied.}
\item {\bf if} $V^{\prime}\cap V^{\prime\prime}\ne V$\\
\hbox{\hspace{2em}} {\bf then} $V\leftarrow V^{\prime}\cap
V^{\prime\prime}$\\
\hbox{\hspace{2em}}\phantom{\bf then } {\bf if} $V\ne\emptyset$ {\bf
then} {\bf goto} {\sl step \ref{repeat}} {\bf else} {\sl stop}\\
\hbox{\hspace{2em}} {\bf else} {\sl stop}
\end{enumerate}
\end{quote}
\vfill
The last step illustrates very important and often used rule of
interval calculations: {\em if the result can be obtained on more than
one way --- do so and take the intersection of partial results as the
final one\/}.  Sometimes at this step $V^{\prime}\cap V^{\prime\prime}$
will be empty.  If this ever happens, then we can be sure, that there
are no solutions within the initial box $V$.  This may mean one of two
things:
\begin{itemize}
\item either our data set contains one or more outliers, or
\item our mathematical model ($f$) is inadequate, the theory is
invalidated by present observations.
\end{itemize}

\medskip\noindent
The {\bf box slicing algorithm}, reducing $p$-dimensional initial box
$V$, is given below.  Explicitly shown is the phase called {\em slicing
from the left\/}.  Slicing from the right is obtained using comments
(surrounded by '\slash*' and `*\slash' pair) instead of original text
in lines marked as 2, 5 and 7.  The complete algorithm consists of both
phases, applied in any order.
\vfill
\begin{quote}
\noindent\phantom{1}1:~{\bf for} $j=1$ {\bf to} $p$ {\bf do}\\
\phantom{1}2:~\hbox{\hspace{2em}}$\xi\leftarrow 1$ \hfill \slash* $\xi \leftarrow
	0$ *\slash\\
\phantom{1}3:~\hbox{\hspace{2em}}$k=1$\\
\phantom{1}4:~\hbox{\hspace{2em}}{\bf repeat}\\
\phantom{1}5:~\hbox{\hspace{4em}}$\xi\leftarrow \xi/2$ \hfill \slash*
	$\xi\leftarrow(1+\xi)/2$ *\slash\\
\phantom{1}6:~\hbox{\hspace{4em}}$k\leftarrow k+1$\\
\phantom{1}7:~\hbox{\hspace{4em}}{\sl consider box} $V^{\prime} =a_1\times a_2\times
\cdots\times\strut{\left[\underline{a}_j,
\xi\left(\overline{a}_j-\underline{a}_j\right)\right]}\times\cdots\times a_p$\\
\phantom{ }\hfill \slash* \hbox{\hspace{4em}}{\sl consider box} $V^{\prime} =a_1\times
a_2\times \cdots\times\strut{\left[\xi\left(\overline{a}_j-\underline{a}_j\right),
\overline{a}_j\right]}\times\cdots\times a_p$ *\slash\\
\phantom{1}8:~\hbox{\hspace{4em}}{\sf success} $\leftarrow$ {\bf not} ({\sl all
conditions/inequalities satisfied in} $V^{\prime}$)\\
\phantom{1}9:~\hbox{\hspace{4em}}{\bf if} {\sf success} {\bf then}
$V\leftarrow V\setminus V^{\prime}$\\
10:~\hbox{\hspace{2em}}{\bf until} ({\sf success} {\bf or} $k>M$)\\
11:~{\bf end for}
\end{quote}
\vfill
The number $M$ denotes simply the number of bits in floating point
representations of real numbers used by a given processor/compiler
pair; for example $M=25$ for single precision reals and $M=57$ for
double precision type numbers in PC-compatible computers equipped with
{\tt g77} or {\tt gcc} compiler.  The variable {\sf success} is of type
{\sl boolean}.

\medskip\noindent
It must be noted, that the procedure outlined in this article produces
the interval hull of possible solutions.  Not every point within the
final box $V$ represents possible solution of the problem, but --- on
the other hand --- no other point, outside $V$, is feasible.  For
graphical illustration see for example \cite{toja}.

\medskip\noindent The ideas expressed here are closely related to the
ones described in \cite{e,f}, however they go much further: instead of
producing just the interval version of well known least squares
procedure, like in \cite{mini2002a}, we have developed completely
different approach, much stronger.  There are, of course, some
drawbacks:
\begin{itemize}
\item the correlations between searched parameters are lost, and
\item the relations of our method with the familiar {\em confidence
level\/} and other statistical terms are still to be determined. 
Probably the famous Chebyshev inequality will be the only effective
tool for this purpose.
\end{itemize}
And what are the advantages?  Well, several:
\begin{itemize}
\item no assumptions are made concerning the distributions of
experimental uncertainties, in particular they need not to be gaussian
(Ockham's razor principle at work),
\item the results are always valid, no matter whether the experimental
uncertainties are `small' or not,
\item it is easy to reliably identify outliers in collected data,
\item uncertainties in both variables are handled naturally and easily,
\item more data usually means less wide intervals for the searched
parameters, in full accordance with common sense,
\item possibly no solution will be obtained, if some uncertainties are
underestimated, de\-li\-bera\-te\-ly or otherwise,
\item reliable bounds for searched parameters (their accuracies)
are produced automatically, without the need for additional analysis. 
They are directly and precisely related to input uncertainties.
\end{itemize}

\medskip\noindent
It is interesting to note, that in \cite{toja}\verb+[+b\verb+]+ we have
found an example, when the `most likely' least squares estimates for
$a$ and $b$ are outside the bounds produced by our box slicing
algorithm.

\section{Acknowledgment}
\noindent
This work was done as a part of author's statutory activity in
Institute of Physics, Polish Academy of Sciences.

\section{Historical note}
\noindent{\small
First traces of `interval thinking' might be attributed to Archimedes
from Syracuse, Greece (287--212 b.c.), famous physicist and
mathematician, who found two-sided bounds for the value of a number
$\pi$: $3\frac{10}{71}\le\pi\le 3\frac{10}{70}$ and a method to
successively improve them. More than 2000 years later, the american
mathematician and physicist, Norbert Wiener, published two papers: {\em
A contribution to the theory of relative position\/} (Proc. Cambridge
Philos. Soc. {\bf 17}, 441-449, 1914) and {\em A new theory of
measurements: a~study in the logic of mathematics\/} (Proc. of the
London Math. Soc., {\bf 19}, 181--205, 1921), in which the two
fundamental physical quantities, namely the position and the time
respectively, were given an interval interpretation.  Only after Second
World War more papers on the subject were written.  Here we have,
probably among several others: chapter 2 of the book {\em Linear
Computations\/} by Paul S.~Dwyer (John Wiley \& Sons, Inc., 1951,
chapter {\em Computation with Approximate Numbers\/}) and {\em Theory
of an Interval Algebra and its Application to Numerical Analysis\/} by
Teruo Sunaga, (RAAG Memoirs, {\bf 2}, 29--46, 1958).  Facsimile of
those and other early papers on interval analysis are freely available
in the web at {\tt http://www.cs.utep.edu\slash interval-comp\slash
early.html.} Here we can also find two papers by polish mathematician
Mieczys{\l}aw Warmus: {\em Calculus of Approximations\/} (Bull. Acad.
Pol. Sci. C1. III, vol. IV (5), 253--259, 1956) and {\em Approximations
and Inequalities in the Calculus of Approximations. Classification of
Approximate Numbers\/} (Bull. Acad. Pol. Sci. math. astr. \& phys.,
vol. IX (4), 241--245, 1961).  And, finally, there are two technical
reports from Lockheed Aircraft Corporation, Missiles and Space
Division, Sunnyvale, California, {\em Interval Analysis I\/} by
R.E.~Moore with C.T.~Yang, {\tt LMSD-285875}, dated September 1959, and
{\em Interval Integrals\/} by R.E.~Moore, Wayman Strother and
C.T.~Yang, {\tt LMSD-703073}, dated August 1960.  R.E.~Moore later
developed more systematic studies in this area, with still more
results presented in his Ph.D. Thesis (Stanford, 1962). He also wrote
the first widely available monograph {\em Interval Analysis\/}
(Prentice Hall, Englewood Cliffs, NJ, 1966) on this topic. Almost
nobody was willing to make any progress in this direction until it was
discovered, that the same problem, programmed in the same computer
language, produces sometimes drastically different results when solved
on different machines.  Due to his accomplishments, R.E.~Moore is
regarded as a founding father of interval analysis. He is still (2003)
active. Besides other things, we owe him the proof of convergence of
interval Newton Method ({\em A test for existence of solutions to
nonlinear systems\/}, SIAM J.~Numer. Anal., {\bf 14} (4), 611--615,
1977).  Since that time we observe growing interest into interval
methods, not only within numerical analysis community.
}



\begin{thebibliography}{9}

\bibitem{Intro} R.B.~Kearfott: {\em Interval Computations: Introduction,
Uses, and Resources\/}, Euromath. Bull. {\bf 2} (1), 95 -- 112, 1996,
{\tt http://interval.louisiana.edu\slash preprints\slash survey.ps}

\bibitem{ftp} {\tt http://www.cs.utep.edu\slash interval-comp\slash}

\bibitem{franc} Luc Jaulin, Michel Kieffer, Olivier Didrit and \'Eric
Walter: {\em Applied Interval Analysis\/}, Springer-Verlag London
Limited 2001

\bibitem{Bill} G.~William Walster: {\em Philosophy and Practicalities of
Interval Arithmetic\/}, in {\em Reliability in Computing}, R.~Moore
(ed.), Academic Press: San Diego, California (1988), pp. 309-323.

\bibitem{toja}\verb+[+a\verb+]+~Marek W.~Gutowski: {\em Interval
straight line fitting\/}, {\tt http://arXiv.org\slash abs\slash
math\slash 0108163},\\
\verb+[+b\verb+]+~Marek W.~Gutowski: {\em Prosta dostatecznie gruba\/}
Post\c{e}py Fizyki, {\bf 53} (4), 181--192, 2002,\\ {\tt
http://pupil.ifpan.edu.pl\slash}\verb+~+{\tt postepy\slash
dodatki\slash prosta\slash prosta.pdf} (in polish)

\bibitem{e} Hung T.~Nguyen, Vladik Kreinovich, and Chin-Wang Tao: {\em
Why 95\% and Two Sigma? A Theoretical Justification for an Empirical
Measurement Practice\/}, {\tt http://utep.edu\slash vladik\slash
2000\slash tr0026a.ps.gz}

\bibitem{f} Luc Longpr\'e, William Gasarch, G.~William Walster, and
Vladik Kreinovich: {\em $m$ Solutions Good, $m-1$ Solutions Better\/},
{\tt http://utep.edu\slash vladik\slash 2000\slash tr0040.ps.gz}

\bibitem{mini2002a} Jie Yang and R. Baker Kearfott: {\em Interval Linear
and Nonlinear Regression --- New Paradigms, Implementations, and
Experiments or New Ways of Thinking of Data Fitting\/}, the talk given
at 2002 SIAM Symposium, Toronto, May 22, 2002, {\tt
http://interval.louisiana.edu\slash preprints\slash 2002}\verb+_+{\tt
SIAM}\verb+_+{\tt minisymposium.ps}

\end{thebibliography}
\end{document}